\documentclass[aps,prl,twocolumn,groupedaddress,superscriptaddress,showpacs]{revtex4}
\usepackage{graphicx}          
\usepackage{dcolumn}           
\usepackage{bm}                
\usepackage{color}
\usepackage[normalem]{ulem}

\begin{document}

\title{Rashba split surface states in BiTeBr}

\author{S.~V. Eremeev}
 \affiliation{Institute of Strength Physics and Materials Science, 634021, Tomsk, Russia\\}
 \affiliation{Tomsk State University, 634050, Tomsk, Russia\\}

\author{I.~P. Rusinov}
 \affiliation{Tomsk State University, 634050, Tomsk, Russia\\}

\author{I.~A. Nechaev}
 \affiliation{Tomsk State University, 634050, Tomsk, Russia\\}
 \affiliation{Donostia International Physics Center (DIPC), 20018 San Sebasti\'an/Donostia, Basque Country, Spain\\}

\author{E.~V. Chulkov}
 \affiliation{Donostia International Physics Center (DIPC), 20018 San Sebasti\'an/Donostia, Basque Country, Spain\\}
 \affiliation{Departamento de F\'{\i}sica de Materiales UPV/EHU, Facultad de Ciencias Qu\'{\i}micas, UPV/EHU, Apdo. 1072, 20080 San Sebasti\'an/Donostia, Basque Country, Spain\\}
 \affiliation{Centro de F\'{\i}sica de Materiales CFM - MPC, Centro Mixto CSIC-UPV/EHU, 20080 San Sebasti\'an/Donostia, Basque Country, Spain\\}

\date{\today}

\begin{abstract}
Within density functional theory, we study bulk band structure
and surface states of BiTeBr. We consider both ordered and
disordered phases which differ in atomic order in the Te-Br
sublattice. On the basis of relativistic ab-initio calculations, we
show that the ordered BiTeBr is energetically preferable as compared
with the disordered one. We demonstrate that both Te- and
Br-terminated surfaces of the ordered BiTeBr hold surface states
with a giant spin-orbit splitting. The Te-terminated
surface-state spin splitting has the Rashba-type behavior with the
coupling parameter $\alpha_R\sim 2$ eV\AA.
\end{abstract}

\pacs{73.20.-r, 75.70.Tj}

\maketitle

\section{Introduction}

Nowadays, a controllable manipulation of the electronic spin
degree-of-freedom without recourse to an external magnetic field is
a process in technological demand, since it constitutes the basis of
functionality of spintronics devices \cite{Zutic_2004}. An obvious
candidate for the phenomenon underlying this process is spin-orbit
interaction (SOI) that couples spin and momentum of electrons. In
the case of two-dimensional (2D) geometries (surfaces, asymmetric
quantum wells, etc.), the SOI may result in a spin splitting of
electron states, which has a nature of the so-called Rashba effect
\cite{Rashba_1960_1984}. This splitting can be tuned by an applied
electric field
\cite{Caviglia_PRL_2010,Studer_PRL2009,Grundler_PRL_2000,Datta_1990,Nitta_1997},
which opens a pathway for realizing electric-field spin
manipulation. There are two key operating characteristics here: the
Rashba energy, $E_R$, and the momentum offset of split states,
$k_R$, which together define the Rashba coupling strength as
$\alpha_R=2E_R/k_R$.

In the conventional semiconductor structures, where the Rashba
effect has been revealed for the first time, the parameter
$\alpha_{\rm R}$ is of order of $10^{-1}$ eV\AA\, (see, e.g., Refs.
\cite{Lommer_PRL_1988,Luo_PRB_1990,Nitta_1997}). However, for
room-temperature applications of spintronics devices, it is crucial
to have $\alpha_R$ as large as possible. As a result, over a long
period of time the Rashba effect keeps attracting a great interest,
and many systems with a large Rashba spin splitting have been
discovered. It was found that the $\bar{\Gamma}$ surface state on
Au(111) has a Rashba splitting with $\alpha_R$ that is about five
times larger than that in the semiconductor heterostructures (see,
e.g., Refs.
\cite{LaShell_PRL_1996,Nicolay_PRB_2001,Hoesch_PRB_2004}). A larger
$\alpha_R$ (half as much as that for Au(111)) has been reported for
a surface state at the Bi(111) surface \cite{Koroteev_PRL_2004}. In
seeking a way to tune spin-orbit-splitting of surface states, it was
shown that, for instance, in the case of the Au(111) surface it can
be done with deposition of Ag-atoms \cite{Cercellier_PRB_2004,
Popovich_PRB_2005}. As was demonstrated in Refs.
\cite{Nakagawa_PRB_2007,Ast_PRL_2007,Bihlmayer_PRB_2007,Ast_PRB_2008,
Mirhosseini_PRB_2009,Gierz_PRB_2010,Bentmann_PRB_2011}, a more
effective way to modify spin splitting of surface states is a
surface alloying of heavy elements (Bi, Pb, Sb) on noble-metal
surfaces. In this case, one arrives at a Rashba-type split surface
state with $\alpha_{\rm R}$ that is about one order of magnitude
greater than that in the semiconductor structures.

To go further in possible tuning of spin-orbit-splitting of electron
states, quantum-well states evolving by the confinement of electrons
in ultrathin metal films have been considered. In the presence of
both the surface and the interface with a substrate, a number of
impacts on the splitting doubles. As was reported, e.g., in Ref.
\cite{Mathias_PRL_2010}, a Bi monolayer film on Cu(111) can provide
with spin-orbit-split quantum-well states in the unoccupied
electronic structure, which are characterized by $\alpha_{\rm R}$
similar to that in the surface alloys. However, apart from a large
spin-orbit splitting, for an efficient spintronics application in
the way specified above a semiconductor substrate and the absence of
spin-degenerate carriers in a quite wide energy interval are more
promising.

In the case of semiconductor substrate (e.g., an ultrathin Pb films
on Si(111) \cite{Dil_PRL_2008}) quantum well states show a Rashba
splitting as small as in the semiconductor structures. A large
Rashba spin splitting on a semiconductor substrate can be reached,
for example, by means of a Bi-trimer adlayer on a Si(111) surface
(see also Ref. \cite{Yaji_NatComm_2010}), where the splitting has a
similar origin as in the Bi/Ag(111) surface alloy and a close value
for the parameter $\alpha_{\rm R}$ \cite{Gierz_PRL_2009}.
Nevertheless, the found spin-split 2D states cannot be well
described by a simple Rashba model, where a parabolic dispersion
with a positive effective mass combines with the spin-splitting that
is linear in electron momentum. This motivates an active search of
new materials and a revision of the already known systems with a SOI
that under certain conditions can lead to a technologically
meaningful spin spliting of a free-electron-like state at a
semiconductor surface. In that sense, the reexamining of bismuth
tellurohalides, where a Rashba-type spin splitting of states has
been revealed to be caused by intrinsic inversion asymmetry of bulk
crystal potential \cite{NatMat_BiTeI,Bahramy}, can be considered as
a great advance made recently in the search. Actually, it was shown
that Te-terminated surfaces of BiTeCl and BiTeI possess a giant
spin-orbit splitting of a free-electron-like surface state
\cite{Eremeev_PRL,Crepaldi_PRL,Landolt_PRL,Eremeev_JETPlett}.

The mentioned bismuth tellurohalides have hexagonal crystal
structures \cite{Shevelkov} and are characterized by ionic bonding
with large charge transfer from Bi to halide- and Te-atomic layers.
The crystal structure is built up of alternating hexagonal layers
Te-Bi-I(Cl) stacked along the hexagonal axis. Besides, each three
layers Te-Bi-I(Cl) form a three-layer (TL) block, and the distance
between the blocks is about one and a half times greater than the
interlayer distances within the Te-Bi-I(Cl) TL structure. Such a
three-layered structure breaks the inversion symmetry of bulk
crystal potential, which leads to appearing of the Rashba-type
spin-orbit splitting of the bulk bands \cite{NatMat_BiTeI}. Due to
the layered crystal structure, the bismuth tellurohalide surfaces
can be terminated by Te- or halide-atom-layer. Both these
terminations hold spin-split surface states
\cite{Eremeev_PRL,Crepaldi_PRL,Landolt_PRL,Eremeev_JETPlett}. These
states emerge by splitting off either from the lowest conduction
band (for the Te-termination) or from the uppermost valence band
(for the halide-atom-termination). The splitting off is caused by
changes in potential (decreasing at the Te-terminated surface and
increasing at the halide-atom-terminated surface) within the
near-surface layers \cite{Eremeev_PRL} as compared with the bulk
region, which is a consequence of strong ionicity.

In addition to BiTeCl and BiTeI, the bismuth-tellurohalide group is
known to have one more semiconductor---BiTeBr. It was previously
reported \cite{Shevelkov,Donges} that its layered crystal structure
is a disordered centrosymmetric one, where tellurium and bromine
atoms randomly distributed within two layers adjacent to Bi-atomic
layer \cite{Shevelkov,Donges}. As a consequence, both the bulk and
surface electronic structure of the disordered BiTeBr was never
addressed before. Recently, bulk electronic structure of the ordered
BiTeBr has been calculated and Rashba-type splitting of some bands
has been analyzed \cite{Eremeev_PRL}.

In the present paper, we examine both disordered and ordered phases
of BiTeBr. We model the crystal structure of the ordered phase as
that of BiTeI but with Br instead of I and the lattice parameters
were taken form Ref.~\cite{Shevelkov}, at that atomic positions are
obtained within a structural optimization. On the basis of ab-initio
calculations we show that the ordered structure is energetically
preferable. We demonstrate that both the Te- and Br-terminated surfaces
of the ordered BiTeBr hold the spin-orbit split surface states
emerged by splitting off from the bulk conduction or valence band
like in other bismuth tellurohalides. For the practical use, as in
the case of BiTeCl and BiTeI the Te-terminated surface of BiTeBr is
more suitable than the halide-atom-terminated one, since it holds a
surface state that has a free-electron-like dispersion and a large
Rashba-type spin splitting. At the same time, BiTeBr has an
advantage over BiTeCl and BiTeI. As compared with BiTeCl, the
bromide has a larger Rashba spin splitting of the Te-terminated
surface state and a wider bulk band gap. In contrast to BiTeI, the
surface state is larger split off from the bulk conduction band and
more isotropic.

\section{Computational method}

The structural optimization and electronic band calculations are
performed within the density functional formalism as implemented in
{\sc VASP} \cite{VASP1,VASP2}. We use the all-electron projector
augmented wave (PAW) \cite{PAW1,PAW2} basis sets with the
generalized gradient approximation (GGA) of Perdew, Burke, and
Ernzerhof (PBE) \cite{GGA_PBE} to the exchange correlation (XC)
potential. The Hamiltonian contains the scalar relativistic
corrections, and the SOI was taken into account by the second
variation method \cite{KH}.

To treat the bulk disordered phase effect, we employ two approaches. The first one
is a supercell approach used within {\sc VASP}, where $4\times
4\times 1$ supercell with several configurations for randomly
distributed Te and Br atoms were considered. The second approach is
a virtual crystal approximation (VCA) as implemented in the {\sc
ABINIT} code \cite{abinit1}, where the configuration averaged
potential of a gray atom occupying a site in the Te-Br sublattice is
defined as a mixture $V_{\mathrm{VCA}} = x V_{\mathrm{Br}}+(1-x)
V_{\mathrm{Te}}$ of Br ($V_{\mathrm{Br}}$) and Te
($V_{\mathrm{Te}}$) pseudopotentials with $x=0.5$. We used GGA-PBE
Hartwigsen-Goedecker-Hutter (HGH) \cite{HGH} relativistic norm-conserving
pseudopotentials taken from Ref. \cite{PSP_HGH} which include the SOI.

The surface of the ordered BiTeBr formed under cleavage can have
Te-layer or Br-layer termination. To simulate semi-infinite
BiTeBr(0001), using {\sc VASP} we consider a 24 atomic layer slab
with bromine side (for Te-terminated surface) or tellurium side (for
Br-terminated surface) passivated by hydrogen monolayer.

\section{Calculation results and discussion}

\begin{figure}
\begin{center}
\includegraphics[width=\columnwidth]{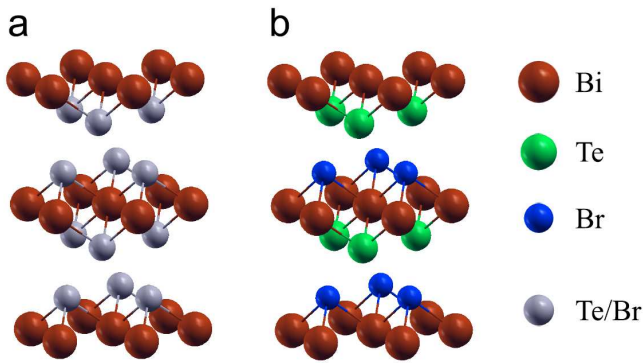}
\end{center}
\caption{Atomic structure of BiTeBr: disordered structure as taken
from Ref.~\cite{Shevelkov} (a) and optimized ordered structure (b).}
\label{geom}
\end{figure}

\begin{figure}
\begin{center}
\includegraphics[width=\columnwidth]{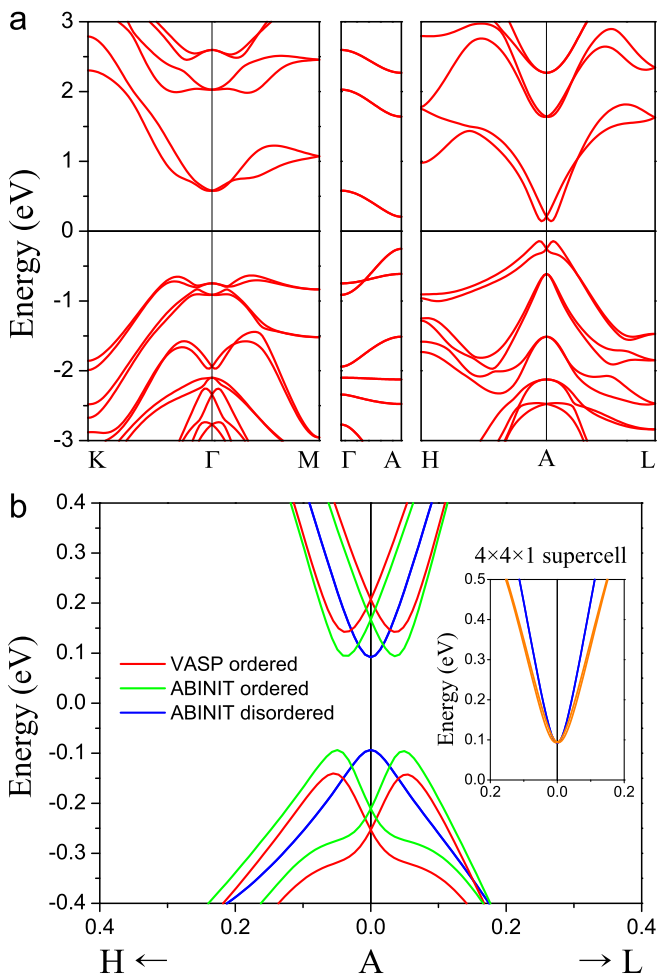}
\end{center}
 \caption{(a) Band structure calculated by the VASP
along high symmetry directions of the Brillouin zone for the BiTeBr
ordered phase, and (b) magnified view of the bulk electronic
structure in the vicinity of the A point calculated with use both
PAW (VASP) and pseudopotential (ABINIT) approaches; inset in
panel (b) shows the lowest conduction band in the vicinity of the A point
calculated within the $4\times 4\times 1$ supercell (orange line) and within the VCA
(blue line).}
 \label{bulk}
\end{figure}

In earlier works \cite{Shevelkov,Donges}, the CdI$_2$ hexagonal
structure for BiTeBr was reported. This structure differs from that
of BiTeI in that Br and Te atoms are statistically distributed over
I$_2$-type sites (Fig.~\ref{geom}(a)). According to
Ref.~\cite{Shevelkov}, the mixed Te/Br layers are located at a
distance of $\pm$1.806 \AA\ from central Bi layer. To study the
geometry of the disordered phase, we constructed a set of $4\times
4\times 1$ supercells with different configurations of randomly
distributed Te and Br atoms. The simulation shows that owing to
markedly different charge transfer from Bi atom to Br and Te atoms,
the Bi layer acquires a substantial rippling of $\sim 0.7$ \AA. This
value is larger than the rippling within the Te/Br layer, which
amounts to $\sim 0.4$ \AA. Such a huge corrugation of atomic layers
leads to a remarkable variation (from 1.3 to 2 \AA) of the distance
between Bi atoms of the central rippled layer and atoms of adjacent
Te/Br layer. As a consequence, the resulting atomic structure
differs essentially from that predicted in Ref.~\cite{Shevelkov}.
Next we considered the BiTeI-type ordered structure
(Fig.~\ref{geom}(b)). In this system, interlayer distances obtained
within the structural optimization are 1.769 \AA\ between Bi and Te
layers and 1.917 \AA\ between Bi and Br layers. The ordered BiTeBr
gains an energy of $\sim 180$ meV per formula unit with respect to
the disordered supercells. The VCA calculation confirms the
preference of the ordered phase.

In Fig.~\ref{bulk}(a), we show the bulk band structure calculated by
VASP code for the ordered BiTeBr. As clearly seen in the figure,
both the conduction band minimum (CBM) and the valence band maximum
(VBM) demonstrate a
giant Rashba-type spin splitting in the H-A-L plane. This spin
splitting is characterized by a slightly anisotropic momentum offset
$k_R$ that in A-H and A-L directions is of $\sim 0.05$ and $0.04$
\AA$^{-1}$ for the VBM and the CBM, respectively. The Rashba energy
for the VBM is approximately twice of that for the CBM (111 meV vs 66
meV). As a result, it provides noticeably larger spin-orbit coupling
for the upper valence-band states as compared with the lower
conduction-band states (see Tabl.~\ref{tab1}).

As seen in the Fig.~\ref{bulk}(b), the pseudopotential {\sc ABINIT} calculations
performed for the ordered BiTeBr confirm the large spin-orbit splitting in
the vicinity of the A point. Moreover, we have obtained values for
the Rashba parameters, which are very close to those found with VASP
(see $\alpha_R$ in Tabl.~\ref{tab1}). The bulk band gap evaluated by
ABINIT is about 100 meV smaller than that obtained from the VASP calculations.
The spectrum calculated within the VCA for the disordered phase
shows practically the same band gap and demonstrates the expected
lack of spin-splitting of the bulk bands due to the presence of
inversion symmetry in the disordered structure. Note that the
spin-splitting of the bulk bands obtained within supercell approach
is negligible, and it agrees well with the VCA result
(Fig.~\ref{bulk}(b), inset), which indicates that chosen $4\times
4\times 1$ geometry is well suited for describing the disordered
BiTeBr.

\begin{table}
 \caption{Rashba coupling parameters $\alpha_R$ (eV\AA) for the bulk
valence and conduction bands in the vicinity of the A point in the A-H
and A-L directions. The calculated values for the bulk band gap,
$E_g$, are also presented.}
 \label{tab1}

\begin{center}
\begin{tabular}{c|c|ll|ll}
\hline
            & $E_g$ (meV) & \multicolumn{2}{c|}{valence band} &\multicolumn{2}{c}{conduction band} \\
            &             & A-H              & A-L            & A-H            & A-L \\
\hline
{\sc VASP}  & 283         & 4.33             & 4.36           & 3.52           & 3.57    \\
\hline
{\sc ABINIT}& 187        & 4.72              & 4.93           & 3.97           & 4.06    \\

\hline
\end{tabular}
\end{center}

\end{table}

\begin{figure}
\begin{center}
\includegraphics[width=\columnwidth]{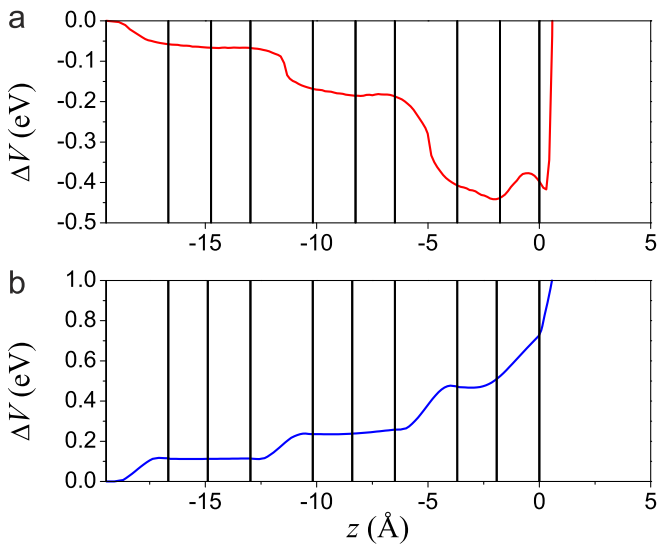}
\end{center}
 \caption{The change of the potential in the near-surface layers of
the crystal with respect to that in the central, bulk-like layers: (a)
Te-terminated surface; (b) Br-terminated surface. $z=0$ corresponds
to the topmost atomic layer.}
 \label{pot}
\end{figure}

As was mentioned above, the spin-split surface states of BiTeI and
BiTeCl emerge as a result of splitting off from the lowest conduction band (at
the Te-terminated surface) or from the uppermost valence band (at the
halide-atom-terminated surface). The splitting off is caused by the
potential change, $\Delta V$, in the near-surface layers of the crystal with respect to that
in central, bulk-like layers. Such a potential change is negative
at the Te-terminated surface and positive at the Cl(I)
atom-terminated surface \cite{Eremeev_PRL,Eremeev_JETPlett}, at that
$\Delta V$ bears a stepwise character owing to clearly defined
three-layered structure of bismuth tellurohalides. A similar
behavior of $\Delta V$ occurs on the surfaces of the ordered BiTeBr,
as seen in Fig.~\ref{pot}, where the change of the potential within the
three outermost TLs on both surface terminations is shown.

\begin{figure*}
\begin{center}
\includegraphics[width=\textwidth]{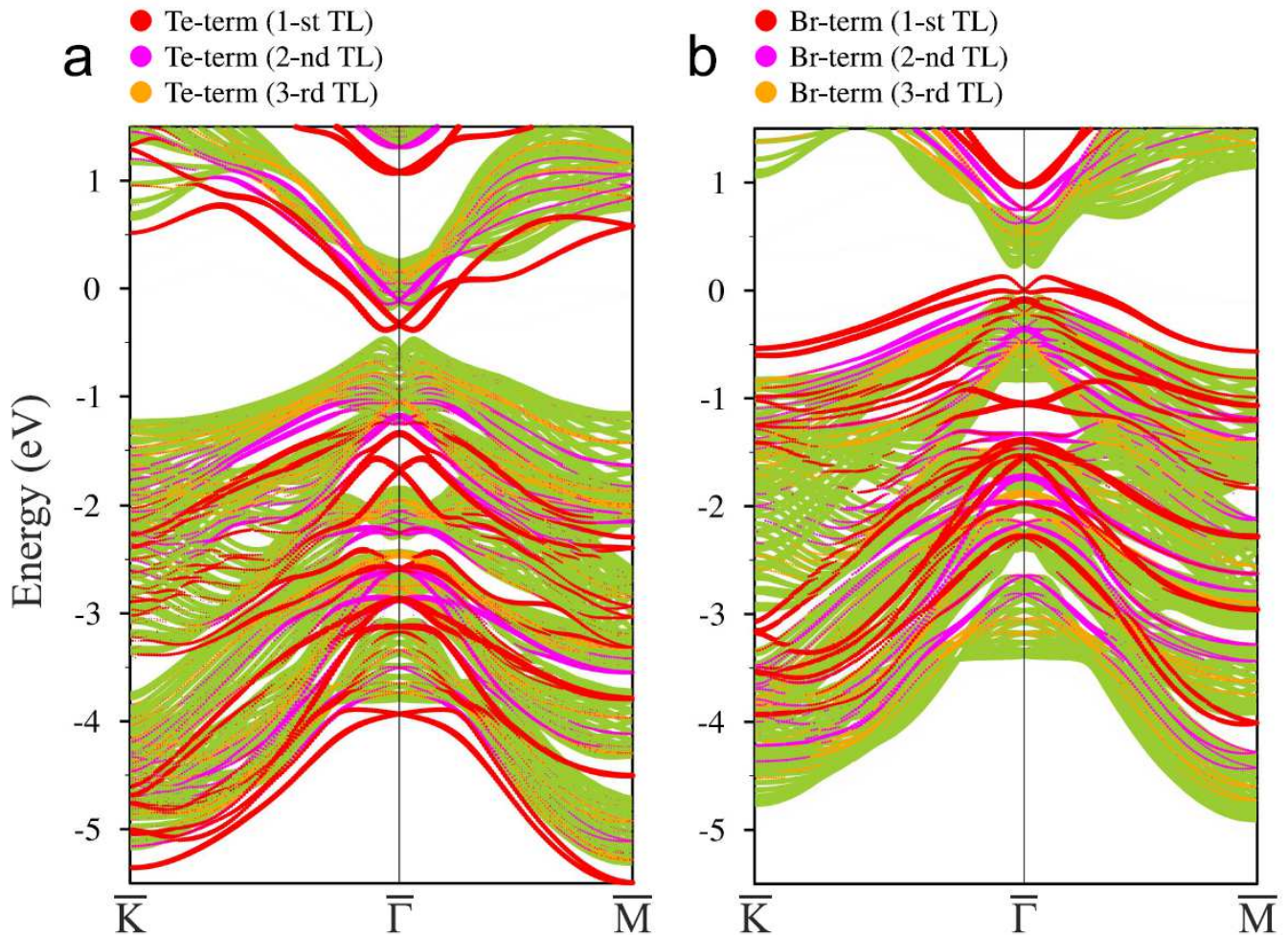}
\end{center}
 \caption{Electronic structure of a BiTeBr(0001) slab: (a)
Te-terminated surface; (b) Br-terminated surface. The red, pink, and
orange circles denote weights of the states localized in the 1-st,
2-nd, and 3-rd TLs of the surface under consideration; light gray
circles mark the states localized on H-terminated side of the slab.
The projected bulk band structure is shown in green.}
 \label{surf}
\end{figure*}

\begin{figure*}
\begin{center}
\includegraphics[width=\textwidth]{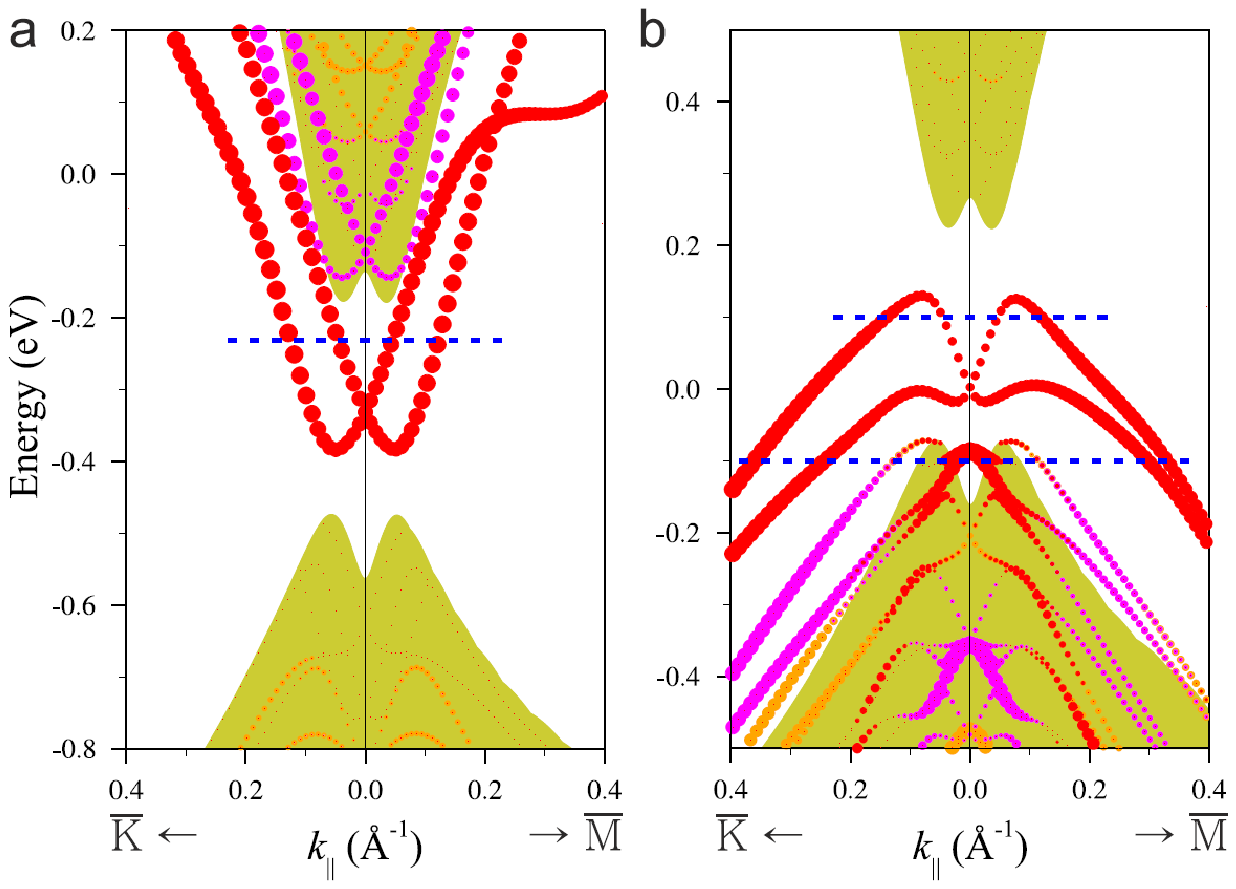}
\end{center}
 \caption{Magnified view of electronic structure of Te-terminated (a) and
Br-terminated (b) BiTeBr(0001) surface in the vicinity of
$\bar\Gamma$ [colors correspond to those marked in
Fig.~\ref{surf}].}
 \label{gap}
\end{figure*}

The negative $\Delta V$ observed at the Te-terminated surface of the
ordered BiTeBr leads to a downward shift of energies of the electron
states trapped in the stepwise surface potential
(Fig.~\ref{surf}(a)). These states are predominantly localized in
the first three TLs. At the Br-terminated surface, an upward shift
of energies of the trapped states is provided by the positive
$\Delta V$ (Fig.~\ref{surf}(b)). The trapped states are offset in
momentum, reflecting a large bulk spin-orbit splitting. They appear
partially overlapping the valence band continuum except the local
energy gap regions within bulk continuum states where the trapped
states can be well resolved in ARPES at high binding energies. As a
net result, for both terminations the changes of the electronic
structure of BiTeBr under the surface formation lead to emergence of
the spin-split surface states in the bulk band gap (Fig.~\ref{gap}).

At the Te-terminated surface (Fig.~\ref{gap}(a)), the spin-orbit
split surface state localized in the topmost TL replicates the
conduction band edge. The degeneracy point of the surface spin-split state
is 150 meV lower than the CBM. Within the energy gap region, the
surface-state dispersion demonstrates the free-electron-like
parabolic character. The spin splitting of the state is
characterized by $\alpha_R = 2.109$ and 2.007 eV\AA\ in the
$\bar\Gamma - \bar{\rm K}$ and $\bar\Gamma - \bar{\rm M}$
directions, respectively.

\begin{figure}
\begin{center}
\includegraphics[width=\columnwidth]{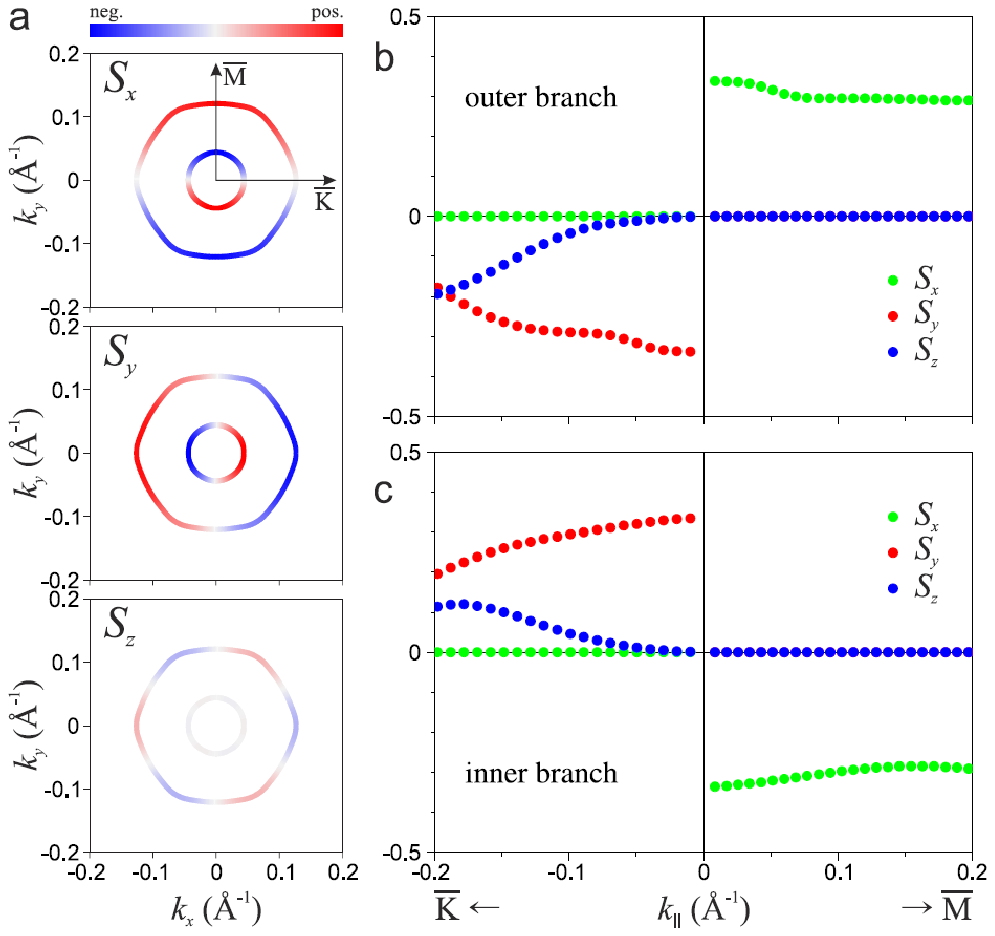}
\end{center}
 \caption{Spin structure of the spin-split surface states on Te-terminated BiTeBr(0001) surface,
as given by spin components $S_x$, $S_y$, and $S_z$ at energy of 100
meV (see horizontal dashed lines in Fig. \ref{surf}) above the degeneracy point (a) and these spin components traced
along the $\bar\Gamma - \bar{\rm K}$ and $\bar\Gamma - \bar{\rm M}$
directions for outer (b) and inner (c) branches of the Rashba-split
surface state. $z$-axis coincides with the surface normal.}
 \label{Te-term_spin}
\end{figure}

The parabolic character of the surface state provides circular shape
of the constant energy contours (CEC) for inner and outer branches
of the spin-split surface state in the bulk band gap region. As one
can see in Fig.~\ref{Te-term_spin}(a), in approaching the bulk
conduction band the CEC for the outer branch acquires the hexagonal
deformation that is already visible at 100 meV above the degeneracy
point. The surface-state spin structure demonstrates
counter-clockwise and clockwise in-plane helicity for the inner and
outer branches, respectively, with a small $S_z$ spin component for
both of them (Fig.~\ref{Te-term_spin}(a)). Owing to symmetry
constrains, the expectation value of the $S_y$ and $S_z$ spin
components vanishes along $\bar\Gamma - \bar{\rm M}$ , and they have
maximal values along $\bar\Gamma - \bar{\rm K}$ at any chosen
energy. In turn, $S_x$ is zero along $\bar\Gamma - \bar{\rm K}$ and
reaches maximal values along $\bar\Gamma - \bar{\rm M}$ direction.
In Fig.~\ref{Te-term_spin}(b), we show the absolute value of the
cartesian spin components as functions of $k_{||}$ for the inner and
outer branches of the spin-split surface state. As one can see, for
a small $k_{\bar{\rm K}}$ the $|S_z|$ component is negligibly small,
and, thus, the surface state is completely in-plane spin polarized.
This component starts rising at $k_{\bar{\rm K}}
> 0.1$ \AA$^{-1}$, i.e. at energy of 50 meV above the degeneracy
point, which leads to a decrease of the in-plane spin components
under approaching the bulk conduction states. Thus, owing to (i) the
parabolic energy dispersion, (ii) outermost TL localization, and
(iii) the in-plane helical spin structure within the band gap energy
region, the surface state on the Te-terminated surface of
BiTeBr(0001) can be described as the Rashba-split surface state with
Rashba coupling parameter of $\sim 2$ eV\AA.

\begin{figure}
\begin{center}
\includegraphics[width=\columnwidth]{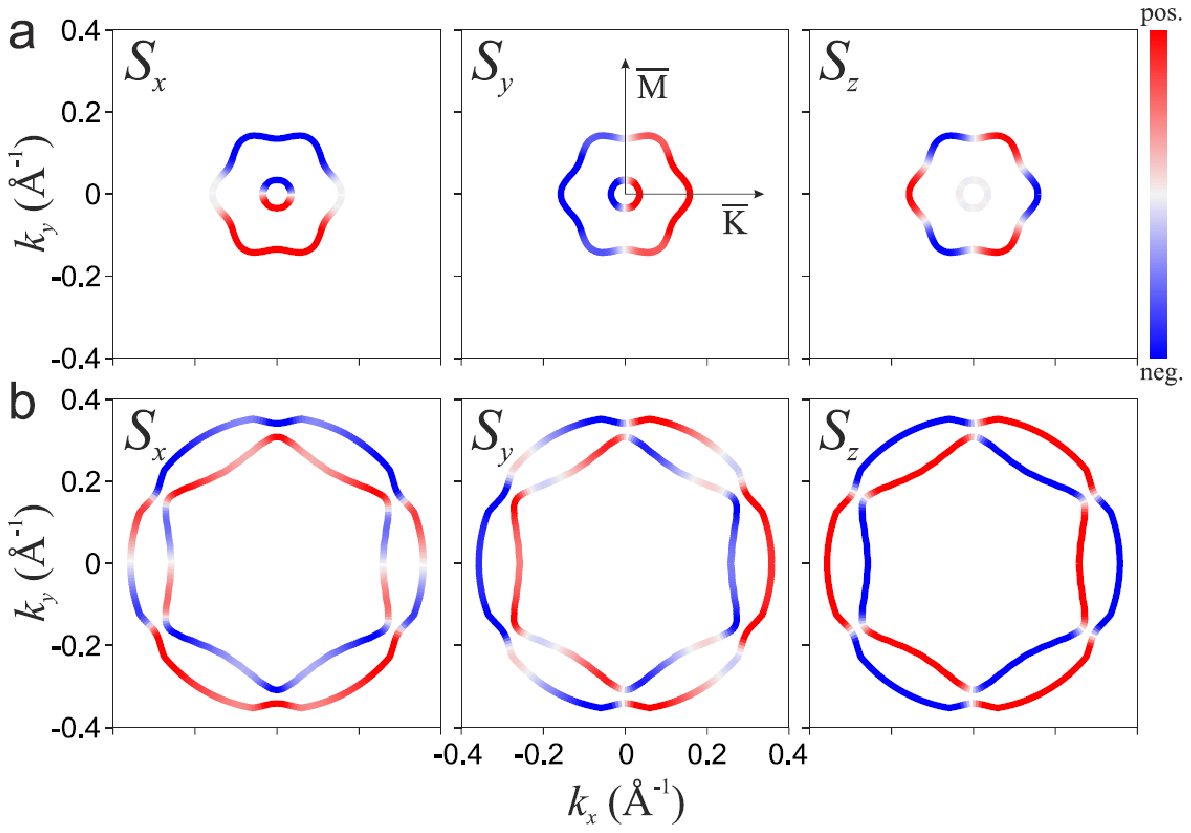}
\end{center}
 \caption{Spin structure of the spin-split surface states on Br-terminated BiTeBr(0001) surface,
as given by spin components $S_x$, $S_y$, and $S_z$ at energy of 100
meV (see horizontal lines in Fig. \ref{surf}) above (a) and below (b) the degeneracy point. $z$-axis coincides
with the surface normal.}
 \label{Br-term_spin}
\end{figure}

At the Br-terminated surface (Fig.~\ref{gap}(b)), the spin-orbit
split surface state localized in the topmost TL lies entirely in the
band gap right above the upper valence band and follows its edge. As
one can see, the states localized in the second TL spread already along the
valence-band edge in $\bar\Gamma - \bar{\rm K}$ and $\bar\Gamma -
\bar{\rm M}$ directions but they degenerate with bulk states in the
close vicinity of $\bar\Gamma$. Furthermore, the second outermost-TL
localized state arises at $\bar\Gamma$ in the valley of the
valence band. The appearance of the second pair of the spin-split
states in the gap and the second state at $\bar\Gamma$ is explained
by the fact that the magnitude of the $\Delta V$ is larger than that
on the Te-terminated surface (see Fig.~\ref{pot}). Such a $\Delta V$
provides a larger splitting off from the valence band edge.

In general features, the band-gap-lying outermost TL-localized
spin-orbit split surface state at the Br-terminated surface
resembles those found at the halide-atom-terminated surface of BiTeCl
and BiTeI \cite{Eremeev_PRL,Eremeev_JETPlett}. This state
demonstrates noticeable anisotropy of the energy dispersion with
respect to $k_{||}$, which results in more complex shape of the CECs
both above and below the degeneracy point (see
Fig.~\ref{Br-term_spin}). Such an anisotropic dispersion is
accompanied by an appreciably out-of-plane spin polarization and
entangled spin structure of the surface state, especially below the
degeneracy point (Fig.~\ref{Br-term_spin}).

Formally, the spin-splitting of the Br-terminated surface state is
characterized by $k_R$ = 0.079 and 0.077 \AA$^{-1}$ in $\bar
\Gamma-\bar{\rm K}$ and $\bar \Gamma-\bar{\rm M}$ direction,
respectively, and by $E_R$ equal to 130.2 meV ($\bar \Gamma-\bar{\rm
K}$) and 125.6 meV ($\bar \Gamma-\bar{\rm M}$). These
characteristics yield $\alpha_R$ equal to 3.29 and 3.26 eV\AA\ for
$\bar \Gamma-\bar{\rm K}$ and $\bar \Gamma-\bar{\rm M}$ directions,
respectively.  However, this spin-split surface state can not be
identified as the Rashba state owing to its dispersion and entangled
spin structure.

\section{Conclusions}

Thus, we have investigated the atomic and electronic structure of
BiTeBr. The total energy calculations of the ordered and disordered
phases of BiTeBr have shown that the ordered structure is
energetically preferable. We have found that the surfaces of the
ordered BiTeBr hold surface states which demonstrate a giant
spin-orbit spin splitting. These states emerge as a result of
splitting off from the bulk conduction or valence band, owing to the
potential bending within the near-surface layers, like in other
bismuth tellurohalides, BiTeCl and BiTeI, studied earlier. The
spin-split surface state at the Te-terminated surface, owing to its
parabolic energy dispersion, outermost TL localization, and in-plane
helical spin structure preserved within the whole band-gap energy
region, can be described as a Rashba-split surface state with the
Rashba coupling parameter $\alpha_R$ of $\sim 2$ eV\AA\, which is in
a good agreement with recently reported experimental value of 2.0(7)
eV\AA \cite{Sakano_arxiv}. The Rashba-split state on Te-terminated
BiTeBr has advantages over the other bismuth tellurohalides, which
consist in the larger Rashba splitting and wider band gap as
compared to BiTeCl and in the larger splitting off from the bulk
conduction band with more isotropic energy dispersion within the
band gap region in comparison with BiTeI.

The band gap surface state at the Br-terminated surface owing
to its $k_{||}$ anisotropy and entangled spin structure can not be
identified as the Rashba-split state and thus has less appeal than
the spin split surface state at the Te-terminated surface.


\end{document}